\documentclass[12pt]{article}
\usepackage{epsfig}
\usepackage{amsmath}

\begin{document}
\vspace*{-.6in}
\thispagestyle{empty}
\begin{flushright}
hep-th/0211198\\
CALT-68-2413\\
NSF-KITP-02-162\\
PUPT-2058\\
UPR-1021-T
\end{flushright}
\baselineskip = 18pt

\vspace{0.25in} {\LARGE
\begin{center}
Explicit Formulas for Neumann Coefficients in the Plane-Wave
Geometry
\end{center}} \vspace{.25in}

\begin{center}
Yang-Hui He${}^1$, John H. Schwarz${}^2$, Marcus
Spradlin${}^3$, and Anastasia Volovich${}^4$\\
\vspace{0.25in}
\emph{${}^1$The University of Pennsylvania\\ Philadelphia, PA  19104, USA}\\
\vspace{0.25in}
\emph{${}^2$California Institute of Technology\\ Pasadena, CA  91125, USA}\\
\vspace{0.25in}
\emph{${}^3$Princeton University\\ Princeton, NJ 08544,  USA}\\
\vspace{0.25in}
\emph{${}^4$Kavli
Institute for Theoretical Physics\\ Santa Barbara, CA  93106, USA}
\end{center}
\vspace{0.25in}

\begin{center}
\textbf{Abstract}
\end{center}
\begin{quotation}
\noindent
We obtain explicit formulas for the Neumann coefficients and
associated quantities that appear in the three-string vertex
for type IIB string theory in a
plane-wave background, for any value of the mass parameter
$\mu$.  The derivation involves constructing the inverse of
a certain infinite-dimensional matrix, in terms of which the
Neumann coefficients previously had been written only implicitly.
We derive
asymptotic expansions for large $\mu$ and find unexpectedly
simple results, which are valid to all orders in $1/\mu$.
Using BMN duality, these give predictions for certain
gauge theory quantities to all orders in 
the modified 't Hooft coupling $\lambda'$. A specific example
is presented.
\end{quotation}

\newpage

\thispagestyle{empty}

\tableofcontents

\pagenumbering{arabic}

\newpage

\section{Introduction}

This paper continues the study of the light-cone gauge string field
theory formulation of type IIB superstring theory in the maximally
supersymmetric plane-wave geometry
\cite{Spradlin:2002ar}-\cite{Pankiewicz:2002tg}.
Its purpose is to derive and analyze explicit
formulas for the Neumann coefficients that enter in the
three-string interaction vertex that describes the process
in which strings \#1 and \#2 join to form string \#3.
In particular \cite{Spradlin:2002ar}, this requires an
explicit formula for the inverse of a certain infinite-dimensional
matrix called $\Gamma_+ (\mu, y)$ (and defined below). Here, $\mu$ is the mass
parameter that appears in the plane-wave metric.  It becomes
physically meaningful once we specify the coordinate frame,
since only the product  $P_- \mu$ is 
invariant under a longitudinal boost. The combination
$\alpha^{\prime} P_-$ for the $r$th string is conventionally denoted
$\alpha_r$, with the momentum taken to be negative for the outgoing string 
so that $\sum_r \alpha_r =0$. We make $\mu$ a meaningful dimensionless
parameter by choosing a frame for which
$\alpha_3 = -1$. When strings \#1 and \#2 join
to form string \#3,
string \#1 carries momentum fraction $\alpha_1 = y$ and 
string \#2 carries
momentum fraction $\alpha_2 = 1-y$, with $0 \le y \le 1$.

\setcounter{page}{1}

Previously \cite{Schwarz:2002bc,Pankiewicz:2002gs}, 
the inverse of $\Gamma_+$ was determined in terms
of a certain infinite component vector called $Y_m (\mu,y) =
(\Gamma_+^{-1} B(y))_m $ and a scalar function $k(\mu,y) = B^{\rm T}
\Gamma_+^{-1} B$. Here $B_m(y)$ is a known infinite component
vector. (The formula will be given later.) In this paper we obtain
explicit formulas for $Y_m$ and $k$, and hence also for
$\Gamma_+^{-1}$. The first step is to derive a first-order
differential equation that determines the dependence of $Y_m$ on
the mass parameter $\mu$. Since the value of $Y_m$ for $\mu =0$ is
known from previous analysis of the flat-space problem, that
knowledge can be used to fix the `initial condition'. The
resulting integrated expression for $Y_m$ is expressed in terms of
$k(\mu, y)$, which still needs to be
determined. Knowledge of the leading large $\mu$ asymptotic
behavior of $Y_m$, for all $m$, is sufficient to completely
determine that function. The resulting equation involves a certain
integral transform. The equation is solved by the inverse integral
transform. The particular integral transform that appears does not seem to be
contained in the standard mathematical references (such as
Bateman), but we have not done an exhaustive search of the literature.

The formulas obtained at this point in the analysis are explicit and
complete,
but they are not yet in a form that is convenient for exploring
large $\mu$ expansions.  The large $\mu$ limit is of
particular interest in light of the
proposed correspondence \cite{Berenstein:2002jq} between type IIB
superstring theory in the plane wave background and a certain
sector of ${\cal{N}} = 4$ SU(N) gauge theory, since the dual gauge
theory is believed to be effectively perturbative in the parameter
$\lambda' = 1/\mu^2$.
In the final section of this paper we present explicit formulas, which
are valid to all orders in $\lambda'$ perturbation theory, but
omit non-perturbative terms of order $e^{-2 \pi \mu |\alpha_r|}$.  
The latter terms can in principle be extracted from our results with
sufficient effort. We use these formulas to make a
specific gauge theory prediction to all orders in $\lambda'.$

\section{Review of Basic Formulas}

The three-string interaction vertex for type IIB superstrings in
flat space was worked out in \cite{Green:1982tc,Green:hw}
and generalized to the plane-wave geometry in
\cite{Spradlin:2002ar,Spradlin:2002rv}.
This vertex is represented as a state in the tensor product
of three string Fock spaces, where the individual strings are
labelled by an index $r=1,2,3$. Any particular three string
coupling is then obtained by contracting it with three specific
string states. The formula for the three-string interaction vertex
contains a bosonic factor
\begin{equation}
\label{vbdef}
|V_B\rangle  = {\rm exp} \left( \frac{1}{2} \sum_{r,s =1}^3 \sum_{m,n
=-\infty}^{\infty} \sum_{I=1}^8 a_{rm}^{I\dagger} \overline{N}_{mn}^{rs}
a_{sn}^{I\dagger}\right) | 0\rangle.
\end{equation}
The quantities $\overline{N}_{mn}^{rs}$ are called Neumann coefficients.
The three-string vertex also contains a similar fermionic factor
$| V_F\rangle$ made out of the fermionic oscillators and a `prefactor'
that is polynomial in the various oscillators. We will not discuss
either of these in this paper. However, aside from an overall factor $v(\mu,y)$
that does not involve the oscillators, they are constructed out
of essentially the same quantities, so the formulas that will be
derived here determine them also.

In describing the Neumann matrices, it is convenient to consider
separately the cases in which each of the indices $m,n$ are either
positive, negative or zero. Henceforth, the symbols $m,n$ will
always denote positive integers. One result of
\cite{Spradlin:2002ar}, for example, using matrix notation for the
blocks with positive indices, is
\begin{equation}
\label{neumann}
\overline{N}^{rs} = \delta^{rs} - 2 (C_r C^{-1})^{1/2} A^{(r){\rm T}}
\Gamma_+^{-1}A^{(s)} (C_s C^{-1})^{1/2}.
\end{equation}
Here $C_{mn} = m \delta_{mn}$ and $(C_r)_{mn} = \omega_{rm}
\delta_{mn}$  are diagonal matrices, with
\begin{equation}
\omega_{rm} = \sqrt{ m^2 + (\mu\alpha_r)^2}.
\end{equation}
The definitions of the matrices $A^{(r)}$ and $\Gamma_+$ will be given
shortly. 

The blocks with both indices negative are related in a simple way
to the ones with both indices positive by
\begin{equation}
\overline{N}_{-m-n}^{rs} = - \left( U_r \overline{N}^{rs} U_s \right)_{mn},
\end{equation}
where
\begin{equation}
U_r = C^{-1} (C_r - \mu \alpha_r) = C (C_r + \mu \alpha_r)^{-1}.
\end{equation}
In the case of $\overline{N}^{33}$ these are the only nonvanishing
terms.  For the remaining Neumann coefficients the other nonvanishing
terms are
\begin{eqnarray}
&&\overline{N}_{m0}^{rs}=\overline{N}_{0m}^{sr} = \sqrt{2\mu\alpha_s}\,
\epsilon^{st}\alpha_t \left[ \left(C_r C^{-1}\right)^{1/2}
A^{(r){\rm T}} \Gamma_+^{-1} B \right]_m, \qquad s,t \in \{ 1,2 \},
r \in \{1,2,3 \},\cr
&&\overline{N}_{00}^{rs}=
(-1)^{r+s+1} (1 + \mu \alpha k)
\frac{\alpha}{\sqrt{\alpha_r \alpha_s}},
 \qquad r,s \in \{ 1,2 \},\cr
&&\overline{N}_{00}^{3r}=\overline{N}_{00}^{r3} = - \sqrt{{\alpha_r}},
\qquad r \in \{ 1,2 \},
\end{eqnarray}
where $\alpha = \alpha_1 \alpha_2 \alpha_3 = - y(1-y)$,
and $k = k(\mu,y)$ will be defined shortly.

The matrices $A^{(r)}(y)$ and the vector $B(y)$ do not depend on
$\mu$, so they are the same as in flat space. Namely,
\begin{eqnarray}
&&A_{mn}^{(1)} = \frac{2}{\pi} (-1)^{m+n+1} \sqrt{mn} \frac{y \sin m
\pi y}{n^2 - m^2 y^2},\cr
&&A_{mn}^{(2)}\! = \!\frac{2}{\pi} (-1)^{m} \sqrt{mn} \frac{(1-y)
\sin m \pi y}{n^2 - m^2 (1-y)^2},\cr
&&A_{mn}^{(3)} = \delta_{mn},\cr
&&B_{m} = \frac{2}{\pi} (-1)^{m +1} \frac{\sin m \pi y}{y(1-y)
m^{3/2}}.
\end{eqnarray}
A number of useful identities relating these matrices
are included in appendix A.
Out of these and the diagonal matrices $U_r$ we construct
\begin{equation}
\Gamma_+ = \sum_{r=1}^3 A^{(r)} U_r A^{(r){\rm T}}.
\end{equation}
Note that the only $\mu$ dependence enters via $U_r$.

The quantities that we especially would like to evaluate
explicitly are the matrix $\Gamma_+^{-1}(\mu, y)$, the vector
\begin{equation}
Y_m(\mu , y) =\left(\Gamma_+^{-1}(\mu , y) B(y)\right)_m,
\end{equation}
and the scalar
\begin{equation}
k(\mu , y) = B^{\rm T} \Gamma_+^{-1} B.
\end{equation}
In the case of flat space ($\mu
=0$) the results are known. Specifically, the Neumann
matrices in eq.~(\ref{neumann}) may be written as
\begin{equation} \label{factorize}
\overline{N}_{mn}^{rs} = - \frac{mn\alpha}{m\alpha_s + n \alpha_r}
\overline{N}_m^r \overline{N}_n^s \qquad {\rm for}\,  \mu =0,
\end{equation}
where
\begin{equation} \label{Nrflat}
\overline{N}_m^r = \frac{\sqrt{m}}{\alpha_r} f_m(-\alpha_{r+1}/\alpha_r)
e^{m\tau_0 /\alpha_r} \qquad {\rm for}\,  \mu =0,
\end{equation}
where $\alpha_4 = \alpha_1$ is understood,
\begin{equation}
f_m(\gamma) = \frac{\Gamma(m\gamma)}{m! \Gamma (m\gamma +1 -m)}
\end{equation}
and
\begin{equation}
\tau_0 = \sum_{r=1}^3 \alpha_r \ln |\alpha_r| = y \ln y +
(1-y) \ln (1-y).
\end{equation}
In particular, still for $\mu =0$, $\Gamma_+^{-1} = \frac{1}{2} (1
- \overline{N}^{33})$, $Y_m = - \overline{N}_m^3$, and $k = 2\tau_0/\alpha$.
In other words,
\begin{equation}\label{Ymzero}
Y_m(\mu = 0,y) = \frac{\sqrt{m}\, \Gamma(my)}{m! \Gamma (my +1
-m)}e^{-m\tau_0} = \frac{\Gamma(1+my) \Gamma(1+m(1-y))}{2\,
\Gamma(1+m)}\, e^{-m\tau_0} B_m
\end{equation}
and
\begin{equation}
k (\mu = 0, y) = - 2 \left(\frac{\ln y}{1 - y} + \frac{\ln
(1-y)}{y}\right).
\end{equation}

In \cite{Schwarz:2002bc,Pankiewicz:2002gs} 
the following identity was derived
for arbitrary $\mu$,
\begin{equation} \label{symformula}
\{ \Gamma_+^{-1} , C_3\} = C + \frac{1}{2} \frac{\alpha_1 \alpha_2}{
1 + \mu \alpha k} C U_3^{-1} Y Y^{\rm T} CU_3^{-1}.
\end{equation}
Note that this determines $\Gamma_+^{-1}(\mu,y)$ in terms of
$Y(\mu,y)$ and $k(\mu,y)$. In particular, this formula was shown to imply that
the generalization of
eq.~(\ref{factorize})  to nonzero $\mu$ takes the form
\begin{equation}
\label{mufactor}
\overline{N}^{rs}_{mn} = - \frac{mn\alpha}{1 + \mu \alpha k}
\frac{\overline{N}_m^r
\overline{N}_n^s}{\alpha_s\omega_{rm} + \alpha_r\omega_{sn}}
\end{equation}
where
\begin{equation}
\label{nvectordef}
\overline{N}_m^r = -\left[ (C^{-1}C_r)^{1/2} U_r^{-1} A^{(r){\rm T}}
Y\right]_m,
\end{equation}
but neither it nor $k(\mu,y)$ was determined explicitly at nonzero $\mu$.

Some preliminary analysis of large $\mu$ asymptotics was initiated in 
\cite{Spradlin:2002rv,Klebanov:2002mp,Schwarz:2002bc,Huang:2002wf}, 
though not much can be done without
additional explicit formulas. It was found that the leading (large
$\mu$) term in the expansion of $\Gamma_+^{-1}$ is given by the
first term on the right-hand side of eq.~(\ref{symformula}).
Defining $R$ by
\begin{equation} \label{decomp}
\Gamma_+^{-1} = \frac{1}{2} C C_3^{-1} + R,
\end{equation}
it is easy to see that
the leading term in $R$ is of order $\mu^{-4}$. Specifically,
\begin{equation}
R \to a_R \pi   \frac{y^2 (1-y)^2
}{\mu^4} {C^3}B B^{\rm T} {C^3} + \cdots
\end{equation}
where $a_R$ is a constant and the next term in the expansion is of
order $\mu^{-6}$. Similarly,
\begin{equation}\label{kasym}
k(\mu,y) \to \frac{1}{\mu y(1-y)}  - \frac{a_k}{\pi [\mu
y(1-y)]^2} +\cdots.
\end{equation}
Inserting these expansions into eq.~(\ref{symformula}), one learns
that
\begin{equation}
\label{akar}
a_R a_k = \frac{1}{64}.
\end{equation}
It is very difficult to determine $a_R$ and $a_k$ separately
without additional explicit formulas. The asymptotic expansion of
$Y$ was found to have the structure
\begin{equation}\label{Yasym}
Y_m  \to \frac{1}{\mu} \left[\frac{1}{2} m - \left(\frac{1}{4}
- x\right) \frac{m^3}{\mu^2} + \ldots \right]
B_m.
\end{equation}
The value of $x$ is of particular interest. It was estimated
numerically to be approximately 1/16 in \cite{Klebanov:2002mp}, and
we will show below that this is correct.

\section{The Differential Equation}

This section describes the derivation of a differential equation
involving $Y_m(\mu, y)$ and $k(\mu , y)$. For the benefit of
the reader who would like to skip the details of the derivation,
and move on to the next section, the result is stated here:
\begin{equation} \label{diffeq}
\frac{\partial Y_m}{\partial \mu} = \left[\frac{1}{2}
\frac{\partial F}{\partial \mu} \left(1 -
\frac{\mu}{\omega_m}\right) - \frac{\mu}{\omega_m^2}\right] Y_m ,
\end{equation}
where $\omega_m = \omega_{3m}=\sqrt{m^2 + \mu^2}$ and
\begin{equation}
\label{Fdef}
F (\mu, y) = \ln [1 + \mu \alpha k (\mu, y)] =
\ln [1 - \mu y (1-y) k (\mu, y)].
\end{equation}
This has the formal solution
\begin{equation}
\label{ysoln}
Y_m (\mu ,y) = \frac{m}{\omega_m} \exp \left[\frac{1}{2}
\int_0^{\mu} \frac{\partial F}{\partial \mu} \left(1 - \frac{\mu}
{\omega_m}\right)d\mu\right] Y_m (0, y).
\end{equation}
Thus, if we knew $k(\mu, y)$, we would know $F(\mu,y)$,
and then $Y_m (\mu, y)$, and
hence all the Neumann coefficients.

The derivation of eq.~(\ref{diffeq}) is rather involved. Let us
sketch the derivation here and then fill in some of the details
in appendix B. The matrix $\Gamma_+ = \sum A^{(r)} U_r A^{(r){\rm T}}$ only
depends on $\mu$ through the dependence of $U_r$ on $\mu$. Its
derivative can be written in the form
\begin{equation}
\frac{\partial \Gamma_+}{\partial \mu} = - \frac{1}{2} \alpha B
B^{\rm T}  + \mu N,
\end{equation}
where
\begin{equation}
N = \sum_{r=1}^3  \alpha_r^2 A^{(r)} C^{-1} C_r^{-1} A^{(r){\rm T}}.
\end{equation}
It follows that
\begin{equation}
\label{follows}
\frac{\partial Y}{\partial \mu} =  \frac{1}{2} k \alpha Y  - \mu
\Gamma_+^{-1} NY.
\end{equation}
But $NY$ can be recast in the form
\begin{equation}
\label{recast}
 NY = g_1 C_3^{-2} B + g_2 B,
\end{equation}
where the coefficients $g_1$ and $g_2$ are scalar quantities:
\begin{eqnarray}
\label{gdefs}
&&g_1 = \frac{2(1 + \mu \alpha k)}{2 + \mu \alpha k + \mu^2 \alpha
k_1},\cr
\cr
&&g_2 = \left(\frac{\alpha}{2}\right) \frac{\alpha k^2  + \mu \alpha k
k_1 + 2 k_1}{2 + \mu \alpha k + \mu^2 \alpha k_1},
\label{gonegtwo}
\end{eqnarray}
and
\begin{equation}
k_i = B^{\rm T} C_3 ^{-i} Y.
\label{ki}
\end{equation}

The above equations imply that
\begin{equation}\label{keq}
\frac{\partial k}{\partial \mu} = B^{\rm T} \frac{\partial Y}{\partial
\mu} = \frac{1}{2} \alpha k^2 - \mu g_2 k -\mu g_1 k_2.
\end{equation}
This is not very useful as it stands, since there is no other
apparent way to determine $k_2$. ($k_1$ could be determined, but
that will turn out not to be necessary.) Substituting the equation
for $NY$ and an identity for $[C_3^{-2}, \Gamma_+^{-1}]$
deduced from eq.~(\ref{symformula}), one can recast the derivative
of $Y$ in the form
\begin{equation}\label{Yeq}
\frac{\partial Y}{\partial \mu} =  (F_0 +  F_1 C_3^{-1} + F_2
C_3^{-2}) Y,
\end{equation}
where the scalar functions $F_i$ are given by
\begin{eqnarray}
&&F_0 = \frac{1}{2} \alpha k - \mu g_2 + \frac{1}{2} \mu g_1 \frac{\alpha
}{1 + \mu \alpha k} (k_1 - \mu k_2),
\cr
&&F_1 = -\frac{1}{2}  \mu g_1 \frac{\alpha}{1 + \mu \alpha k} (k -
\mu^2 k_2),
\cr
&&F_2 =  - \mu g_1 + \frac{1}{2} \mu^2 g_1 \frac{\alpha}{1 + \mu
\alpha k} (k - \mu k_1).
\label{expf}
\end{eqnarray}
Using eq.~(\ref{gdefs}) for $g_1$ and $g_2$, and eq.~(\ref{keq})
to eliminate $k_2$ in favor of $k'$, we find
\begin{equation}
F_2 = - \mu, \qquad F_1 = - \mu F_0, \qquad F_0 =
\frac{\alpha}{2} \frac{1}{1 + \mu \alpha k} ( k + \mu k'),
\end{equation}
so that eq.~(\ref{Yeq}) can be written in the form (\ref{diffeq}).

As a first application of eq.~(\ref{diffeq}), let us consider it
for large $\mu$. The expansion in eq.~(\ref{kasym}) implies that
at large $\mu$ the leading behavior of $\partial F/ \partial \mu$
is $-1/\mu$. Substituting this into eq.~(\ref{Yasym}), one deduces that
$x= 1/16$.

\section{An Integral Transform}

In this section we show that the known large $\mu$ behavior
$Y_m \sim \frac{m}{2 \mu} B_m$,
together with eq.~(\ref{ysoln}), is sufficient to deduce
an integral transform, whose solution we
are able to determine explicitly, thereby obtaining
explicit formulas for $k(\mu,y)$, $F(\mu,y)$ and hence
$Y_m(\mu, y)$.

Since $Y_m \sim \frac{m}{2\mu} B_m$ for large $\mu$, 
eq.~(\ref{ysoln}) implies that
\begin{equation}
\label{thirtyeight}
\exp \left\{ \frac{1}{2} \int_0^\infty \frac{\partial F}{\partial
\mu} \left( 1 - \frac{\mu}{\omega_m}\right) d\mu \right\} =
\frac{B_m}{2Y_m(0)}.
\end{equation}
Taking the logarithm of both sides and integrating by parts, this
becomes
\begin{equation}\label{intF}
\int_0^\infty (m^2 + \mu^2)^{-3/2} F(\mu, y) d\mu = G(m , y),
\end{equation}
where, using eq.~(\ref{Ymzero}),
\begin{equation}
\label{Gdef}
G(z, y) = \frac{2\tau_0}{z} + \frac{2}{z^2} \ln \left(
\frac{\Gamma(1+z)}{\Gamma( 1+ zy) \Gamma(1+ z(1-y))} \right).
\end{equation}
Eq.~(\ref{intF}), which must hold for $m = 1,2, ...,$ and $0 \leq
y \leq 1$, determines $F(\mu, y)$ and hence $k(\mu, y)$.

The function $G(z,y)$ is holomorphic in the right-half $z$ plane
and goes to 0 as $z\to \infty$ in that half plane like a power
($1/z^2$). These properties make this extrapolation from the
positive integers to a continuous variable $z$ unique. 
Moreover, they are exactly what is
needed to construct an inverse transformation giving $F$ in terms
of $G$. Let us state this as:

\noindent{\bf Theorem.} Suppose that $g(z)$ is holomorphic in the
right-half $z$-plane and vanishes like a power at infinity in that
half plane. Then the solution of the equation
\begin{equation}\label{gis}
\int_0^\infty (z^2 + x^2)^{-3/2} f(x) dx = g(z)
\end{equation}
(for ${\rm Re}(z)>0$)
is given by the inverse integral transform
\begin{equation} \label{fis}
f(x) = -i \frac{x^2}{\pi} \int_0^\pi  g(-ix {\rm cos} \,\theta) \,
{\rm cos} \, \theta\, d \theta = -i \frac{x^2}{\pi} \int_0^{\pi/2}  
\left[ g(-ix {\rm cos} \,\theta) \,
 -  g(ix {\rm cos} \,\theta)\right] \,
{\rm cos}\, \theta\, d \theta.
\end{equation}
We refer the reader to appendix C for the proof of this theorem.

Applying the theorem to the problem at hand, we learn that
\begin{equation} \label{Fis}
F(\mu, y) = -i \frac{{\mu}^2}{\pi} \int_0^\pi  G(-i\mu \cos\theta,y)
\cos\theta\, d \theta .
\end{equation}
Substituting the formula for $G$,
\begin{equation}
F(\mu, y) = 2\mu\tau_0 + \frac{2i}{\pi} \int_0^\pi  \frac{ d \theta
}{\cos \theta}\, \ln \left[ \frac{\Gamma(1-i\mu \cos \theta)}{
\Gamma(1-i\mu y \cos\theta)
\Gamma(1-i\mu (1-y) \cos \theta)} \right].
\end{equation}
The formula for $Y_m$ also requires the derivative
\begin{eqnarray}
&&\frac{\partial F(\mu, y)}{\partial \mu} = 2\tau_0 + \frac{2}{\pi}
\int_0^\pi  d \theta   \big[ \psi(1-i\mu \cos\theta)
-y \psi(1-i\mu y \cos\theta)
\cr
&&\qquad\qquad\qquad
\qquad\qquad\qquad\qquad\qquad-
(1-y) \psi(1-i\mu (1-y)\cos \theta) \big],
\end{eqnarray}
where, as usual, $\psi = (\ln \Gamma)'$.

Substituting the expansion
\begin{equation}
\psi(1+z) = -\gamma + z \sum_{n=1}^{\infty} \frac{1}{n(z+n)}
\end{equation}
and using the integral
\begin{equation}
\int_0^{\pi} \frac{d \theta}{a + b\cos\theta } = \frac{\pi
}{\sqrt{a^2 - b^2}}
\end{equation}
gives the result
\begin{equation}
\frac{\partial F(\mu, y)}{\partial \mu} = 2\tau_0 + 2
\sum_{n=1}^{\infty} \sum_{r=1}^3 \frac{\alpha_r}{\omega_{rn}}.
\end{equation}
Since $F(0, y) = 0$, this integrates to
\begin{equation}
 F(\mu, y) = 2\tau_0 \mu + 2
\sum_{n,\, r} \ln \left[( \omega_{rn}+ \mu\alpha_r)/n
\right] = 2\tau_0 \mu - 2\, \ln\det(U_1 U_2 U_3) .
\end{equation}
Note that $\det(U_1 U_2 U_3)$ is convergent even though the
individual $\det(U_r)$ diverge. We can regulate them in a
way that does not change the product (since $\sum \alpha_r = 0$)
by recasting $F$ in the form
\begin{equation}
\label{Fresult}
F(\mu, y) = 2\tau_0 \mu + 2 \sum_{r} \phi(\mu\alpha_r) ,
\end{equation}
where
\begin{equation}
\label{phidef}
\phi(x) =  \sum_{n=1}^{\infty}\left[ \ln \left( \frac{\sqrt{n^2 +
x^2} + x}{n} \right) -\frac{x}{n}\right] =
\sum_{n=1}^{\infty}\left[ {\rm arc~sinh}\left(\frac{x}{n} \right)
-\frac{x}{n}\right].
\end{equation}
The regulated determinant of $U_r$ is then ${\rm exp}[
-\phi(\mu\alpha_r)]$. Substituting the expansion of $\partial
F/\partial \mu$ into the formula for $Y_m$ gives
\begin{equation}
\label{ymresult}
Y_m (\mu ,y) = {\rm exp}\left[(\mu - \omega_m)\tau_0
+\sum_{r=1}^3(\phi_r - \phi_{mr})\right]~ \frac{m}{2\omega_m} B_m,
\end{equation}
where $\phi_r = \phi(\mu \alpha_r)$ and
\begin{equation}
\label{phimrdef}
\phi_{mr} = \sum_{n=1}^{\infty} \left[ \ln\left(  \frac{
\omega_{rn} +\omega_m \alpha_r}{n} \right) - \frac{\omega_m
\alpha_r}{n} \right],
\end{equation}
where, we remind the reader, $\omega_m = \sqrt{m^2 + \mu^2}$ and
$\omega_{rm} = \sqrt{ m^2 + (\mu\alpha_r)^2}$.

The formulas (\ref{Fresult}) and (\ref{ymresult})
are explicit expansions of $F$ and $Y_m$
which converge for all finite $\mu$. Thus, in a sense, they solve
our problem. However, they are not yet in the most convenient form
for exploring large $\mu$ expansions. 

\section{Asymptotic Expansions for Large $\mu$}

In this section we develop asymptotic
large $\mu$ expansions for all quantities that appear in the
three-string vertex.
It turns out that the quantities all have essential singularities at
$\mu = \infty$ arising from terms proportional to
$e^{- 2 \pi \mu |\alpha_r|}$.   The existence of such terms,
which correspond to non-perturbative effects in the dual gauge
theory, has been noted in \cite{Klebanov:2002mp}.
We proceed under the assumption that $\mu |\alpha_r|$
is sufficiently large (for all $r$)
that these terms can be neglected, and we 
use the symbol $\approx$ to denote this approximation.
We stress that
the formulas we present encapsulate all orders in a power
series expansion around $\lambda' = 0$ from the dual gauge
theory point of view.  However, they do not match smoothly
with the known flat space results at $\mu = 0$ because
the omitted terms become important in this limit.
Finally, the limit $y \to 0$ (at fixed $\mu$) is also of interest, because
it can be used to extract
vertex operators
for the emission of on-shell particles \cite{Green:1982tc}.
Our asymptotic formulas are not well-suited for
studying this limit because we omit terms
of order $e^{-2 \pi \mu y}$.

Let us demonstrate how
to use the integral 
equation (\ref{intF}) directly to find
the constant $a_k$ introduced in
eq.~(\ref{kasym}).
It  follows from eq.~(\ref{kasym}) that
\begin{equation}
\label{flargemu}
F(\mu,y) = - \ln \left[ {\pi \mu y (1-y)}/a_k \right] + \cdots
\end{equation}
at large $\mu$.
Let us define $\widetilde{F}(\mu,y) = F(\mu,y) +
\ln \left[ {\pi \mu y (1-y)}/a_k \right]$.  Then
eq.~(\ref{intF}) implies that
\begin{equation}
\int_0^\infty (m^2 + \mu^2)^{-3/2} \widetilde{F}(\mu,y) d\mu =
G(m,y) + \frac{1}{m^2} \ln\left(\frac{ m \pi y (1-y)}{2 a_k}
\right).
\end{equation}
Now let $m = n \lambda$ and $\mu = \lambda w$.  After scaling
out $\lambda$, we find
\begin{equation}
\int_0^\infty  (w^2 + n^2)^{-3/2} \widetilde{F}(\lambda w, y)
dw = \lambda^2 \left[ G(\lambda n, y) + \frac{1}{\lambda^2 n^2}
\ln \left(\frac{\lambda n \pi y (1-y)}{2 a_k} \right)\right].
\end{equation}
In the limit $\lambda \to \infty$ the left-hand side goes to
zero, while from 
eq.~(\ref{Gdef}) it is easy to check using Stirling's
approximation that the right-hand side
goes to
$\frac{1}{n^2} \ln(4 a_k)$.
This determines $a_k = 1/4$ and, from
eq.~(\ref{akar}), $a_R = \frac{1}{16}$.

\subsection{$F(\mu,y)$ and $k(\mu,y)$}

Let us proceed by studying
the function $\phi(x)$ defined in eq.~(\ref{phidef}).
Since $\phi$ is clearly odd, it is sufficient to consider
large positive $x$ here.
Taking two derivatives gives
\begin{equation}
\label{ddphi}
\phi''(x) = - x \sum_{n=1}^\infty \frac{1}{(x^2 + n^2)^{3/2}}
\approx - \frac{1}{x} + \frac{1}{2 x^2},
\end{equation}
where we have used eq.~(\ref{sumoneb}) to evaluate the sum (the
exact result is given in eq.~(\ref{phippexact})).
Integrating twice with respect to $x$
leads to
\begin{equation}
\label{phiresult}
\phi(x) \approx - (x + {\textstyle{ \frac{1}{2} }}) \ln x + c_1 x + c_2,
\end{equation}
where $c_i$ are constants of integration.
Inserting eq.~(\ref{phiresult}) into eq.~(\ref{Fresult}) gives
\begin{equation}
\label{Fexpone}
F(\mu,y) \approx - \ln [ \mu y (1-y)] + 2 c_2.
\end{equation}
The constant $c_1$ has dropped out since $\sum \alpha_r = 0$,
and we can determine $c_2 = \frac{1}{2} \ln 4 \pi$ by comparing
eq.~(\ref{Fexpone}) to eq.~(\ref{flargemu}) with $a_k = \frac{1}{4}$,
obtaining thereby
\begin{equation}
\label{Fexpansion}
F(\mu,y) \approx - \ln [ 4 \pi \mu y (1-y)].
\end{equation}
We emphasize that eq.~(\ref{Fexpansion}) is much stronger
than eq.~(\ref{flargemu}):  when we wrote the latter, we might
have expected corrections involving powers of $1/\mu$, but in
eq.~(\ref{Fexpansion}) we have proven that the only corrections
are exponentially small.
An alternative derivation of this result is presented in appendix
D, where we show that the absence of power law corrections follows
from a simple contour integral argument.  In appendix F we derive the
exact formula
\begin{equation}
\label{fullF}
F(\mu,y) = - \ln [4 \pi \mu y(1-y)] + J(\mu y) + J(\mu(1-y)) - J(\mu),
\end{equation}
where
\begin{equation}
\label{Jdef}
J(x) = \frac{2}{\pi} \int_1^\infty \frac{\ln(1 - e^{-2 \pi x z})
}{z \sqrt{z^2 - 1}} dz.
\end{equation}
It is easy to read off all the exponential corrections to
eq.~(\ref{fullF}) by writing out the series expansion of the logarithm.
However, we will not keep track of these exponential corrections
in the following sections, and instead simply use
eq.~(\ref{Fexpansion}) and the definition~(\ref{Fdef}) to write
\begin{equation}
\label{kexpansion}
k(\mu,y) \approx \frac{1}{\mu y(1-y)} - \frac{1}{4 \pi \mu^2 y^2 (1-y)^2}.
\end{equation}

\subsection{$Y_m(\mu,y)$}

Although it is straightforward to develop an asymptotic expansion
for the function $\phi_{mr}$ defined in eq.~(\ref{phimrdef}),
a more direct route is simply to rewrite eq.~(\ref{ysoln}) as
\begin{equation}
Y_m(\mu,y) = \frac{m}{\omega_m} \exp \left[
\frac{1}{2} \int_0^\infty \frac{\partial F
}{\partial \mu} \left(1 - \frac{\mu}{\omega_m} \right) d\mu
- \int_\mu^\infty \frac{\partial F
}{\partial \mu} \left(1 - \frac{\mu}{\omega_m} \right) d\mu
\right] Y_m(0,y).
\end{equation}
The first integral is just eq.~(\ref{thirtyeight}), and the second
integral is elementary after substituting the asymptotic
expansion $F' \approx - {1/ \mu}$.  The final result is
\begin{equation}
\label{finaly}
Y_m(\mu,y) \approx \sqrt{\frac{\mu+\omega_m}{2\mu}} \frac{m}{2 \omega_m}
B_m = \frac{1}{2 \sqrt{2 \mu}} \frac{ \sqrt{\mu + \sqrt{m^2 + \mu^2}}
}{\sqrt{m^2 + \mu^2}} m B_m,
\end{equation}
which can be conveniently summarized as
\begin{equation}
\label{yapprox}
Y \approx \frac{1}{2 \sqrt{2 \mu}} U_3^{1/2} C^{3/2} C_3^{-1}  B.
\end{equation}

\subsection{The Neumann vectors}

\noindent
The final step in the construction
of the Neumann matrices involves evaluating the matrix
products $A^{(r) {\rm T}} Y$ which appear
in the Neumann vectors in eq.~(\ref{nvectordef}). From
eq.~(\ref{finaly}) we have
\begin{equation}
(A^{(1) {\rm T}} Y)_n
\approx \sqrt{\frac{2}{\mu}}
\frac{(-1)^{n+1} \sqrt{n}}{\pi^2 y^2 (1-y)}
\sum_{m=1}^\infty
\frac{ \sin^2 (\pi m y)}{m^2 - n^2/y^2}
\frac{\sqrt{\mu + \sqrt{\mu^2 + m^2}}}{
\sqrt{\mu^2 + m^2}}.
\end{equation}
Using eq.~(\ref{sumthree}), we find
\begin{equation}
\label{aty}
(A^{(r) {\rm T}} Y)_n \approx
\frac{1}{\pi y (1-y) 2 \sqrt{2 \mu}} (-1)^{r(n+1)-1}  \sqrt{\alpha_r}
\omega_{rn}^{-1}
U_{rn}^{1/2}, \qquad r  \in \{1,2\}.
\end{equation}
The Neumann vectors in eq.~(\ref{nvectordef}) can therefore
be expressed as
\begin{eqnarray}
&&\overline{N}^r_n \approx 
 \frac{1}{2  \pi y(1-y)}
 (-1)^{r(n+1)}  \sqrt{\alpha_r}
(2 \mu n \omega_{rn} U_{rn})^{-1/2}, \qquad r  \in \{1,2 \},\cr
&&\overline{N}^3_n \approx - {\frac{n}{2}}
(2 \mu \omega_{3n} U_{3n})^{-1/2} B_n.
\end{eqnarray}
Actually we can combine these expressions in a useful way.
If we define
\begin{equation}
\label{sdef}
s_{1m} = s_{2m} = 1, \qquad s_{3m} = - 2 \sin(\pi m y),
\end{equation}
then we have simply
\begin{equation}
\label{finalnv}
\overline{N}^r_n \approx \frac{(-1)^{r(n+1)} 
}{2 \pi y(1-y)}
\sqrt{  \frac{|\alpha_r|}{2 \mu n \omega_{rn} U_{rn}}
} s_{rn}, \qquad r \in \{1,2,3 \}.
\end{equation}

\subsection{Consistency checks}

We should check that the vector $Y$ we have found indeed satisfies
$B^{\rm T} Y \approx k$ and $\Gamma_+ Y \approx B$.
First we have
\begin{equation}
B^{\rm T} Y \approx \frac{4}{\pi^2}
\frac{1}{y^2 (1-y)^2} \frac{1}{2 \sqrt{2 \mu}}
\sum_{m=1}^\infty \frac{\sin^2(\pi m y)
}{m^2} \frac{\sqrt{\mu + \sqrt{\mu^2 + m^2}}}{
\sqrt{\mu^2 + m^2}}.
\end{equation}
With the help of eq.~(\ref{sumfour}), it immediately follows
that this expression equals
eq.~(\ref{kexpansion}).

In \cite{Klebanov:2002mp} it was shown that $\Gamma_+ = 2 C_3 C^{-1}
- H$, where
$H$ is given up to exponential corrections by
\begin{equation}
\label{hdef}
H_{mn}  \approx \frac{8}{\mu^2 \pi^2} (-1)^{m+n}
\sqrt{m n} \sin(\pi m y) \sin(\pi n y) \int_1^\infty dz
\frac{ \sqrt{z^2 - 1} }{(z^2 + m^2/\mu^2) (z^2 + n^2/\mu^2)}.
\end{equation}
(The integral is easily evaluated, but it is convenient to
leave eq.~(\ref{hdef}) in this form for the calculation.)
The condition $\Gamma_+ Y = B$ which we would now like
to check is equivalent to
\begin{equation}
\label{tocheck}
H Y = 2 C_3 C^{-1} Y - B.
\end{equation}
Using eqs.~(\ref{hdef}) and~(\ref{finaly}) we can write
\begin{equation}
\label{blahthree}
(H Y)_n \approx \frac{2 \sqrt{2} n^2 B_n }{ \pi^2 \sqrt{\mu}} 
\int_1^\infty \frac{ \sqrt{z^2 - 1} }{ z^2 + n^2/\mu^2}
\sum_{m=1}^\infty \frac{\sin(\pi m y)^2 }{ m^2 + z^2 \mu^2}
\frac{ \sqrt{\mu + \sqrt{\mu^2 + m^2}} }{ \sqrt{m^2 + \mu^2}}.
\end{equation}
After substituting eq.~(\ref{sumfive}) for the sum,
the remaining integral over $z$ is elementary and takes the form
\begin{equation}
\label{blahfour}
{\rm Re}
\int_1^\infty dz \frac{ \sqrt{-1 + i \sqrt{z^2 - 1}}
}{ z (z^2 + a^2)} =
- \frac{\pi}{\sqrt{2} a^2}
\left[ 1 - \frac{1}{\sqrt{2}} \sqrt{1 + \sqrt{1 + a^2}}\right]
\end{equation}
for $a = n/\mu$.
Assembling all factors from eqs.~(\ref{blahthree}),~(\ref{blahfour})
and~(\ref{sumfive}), we find
\begin{equation}
(H Y)_n \approx - B_n \left[
 1 - \frac{1}{\sqrt{2}} \sqrt{1 + \sqrt{1 + n^2/\mu^2}}\right].
\end{equation}
Recalling eq.~(\ref{finaly}), we see that the desired relation
eq.~(\ref{tocheck}) is indeed satisfied up to exponential corrections.

\subsection{Some remaining quantities}

The Neumann matrices are completely determined, to all orders
in $1/\mu$, by the
factorization identity (\ref{mufactor})
and the Neumann vectors given in eq.~(\ref{finalnv}).
For the sake of completeness, we catalog here two more quantities
of interest, which follow easily
from our results. The first is the matrix $R$ defined in eq.~(\ref{decomp}).
By comparing the expressions (\ref{neumann}) and (\ref{mufactor})
for $r=s=3$ we can determine $\Gamma_+^{-1}$, and hence
$R$, in terms of $Y_m$.
Using eq.~(\ref{yapprox}), we arrive at
\begin{equation}
R_{mn} = (\Gamma_+^{-1})_{mn} -
\frac{1}{2} \frac{m}{\sqrt{m^2 + \mu^2}}
\delta_{mn}\approx \frac{1}{\pi} m n (-1)^{m + n} \frac{\sin(\pi m y)
\sin(\pi n y)
}{ \omega_n \omega_m (\omega_m + \omega_n)\sqrt{U_m U_n}},
\end{equation}
where (as before)
\begin{equation}
\omega_m = \sqrt{m^2 + \mu^2}, \qquad U_m = \frac{ \omega_m + \mu }{ m}.
\end{equation}

Finally, we remarked above that the prefactor is polynomial
in quantities $X$ and $Y$ which are constructed
out of bosonic and fermionic oscillators respectively
(see \cite{Spradlin:2002rv,Pankiewicz:2002gs,Pankiewicz:2002tg}).
The normalization of $X$ and $Y$
involves a factor called $f(\mu)$ in \cite{Pankiewicz:2002gs}.
It may be obtained from
the formula \cite{Spradlin:2002rv,Pankiewicz:2002gs}
\begin{equation}
f(\mu) = \sqrt{-2 \pi \alpha_2 \alpha_3} \lim_{n \to \infty}
(-1)^n (C A^{(1) {\rm T}}
\Gamma_+^{-1} B)_n.
\end{equation}
It follows immediately from eq.~(\ref{aty}) that
\begin{equation}
f(\mu) \approx \frac{1}{\sqrt{4 \pi \mu y (1-y)}}
\approx \sqrt{1 + \mu \alpha k}.
\end{equation}
In fact, closure of the
supersymmetry algebra requires $f(\mu) = \sqrt{1 + \mu \alpha k}$
\cite{Pankiewicz:2002tg}.
Note that this $f(\mu)$ is separate from a still undetermined
function $v(\mu)$, to appear below, which appears as an overall
factor in the cubic part of the Hamiltonian and supersymmetry
generators.

\subsection{Summary of Neumann matrices}

We summarize here the final expressions for the Neumann matrices
obtained in this paper.  As before, we use the notation
\begin{equation}
\alpha_1 = y, \qquad \alpha_2 = 1-y, \qquad \alpha_3 = -1,
\end{equation}
\begin{equation}
s_{1m} = s_{2m} = 1, \qquad s_{3m} = - 2 \sin(\pi m y),
\end{equation}
and $\omega_{rm} = \sqrt{m^2 + (\mu \alpha_r)^2}$.
Then for $m,n>0$ we have
\begin{equation}
\label{mainresultone}
\overline{N}^{rs}_{mn} \approx \frac{1}{
2 \pi}\frac{ (-1)^{r(m+1) + s(n+1)} }{
\alpha_s \omega_{rm} + \alpha_r \omega_{sn}}
\sqrt{\frac{ |\alpha_r \alpha_s| (\omega_{rm}
+ \mu \alpha_r) (\omega_{sn} + \mu \alpha_s) }{ \omega_{rm} \omega_{sn}
}} s_{rm} s_{sn},
\end{equation}
\begin{equation}
\label{mainresulttwo}
\overline{N}^{rs}_{-m,-n} \approx - \frac{1}{
2 \pi}\frac{ (-1)^{r(m+1) + s(n+1)} }{
\alpha_s \omega_{rm} + \alpha_r \omega_{sn}}
\sqrt{\frac{ |\alpha_r \alpha_s| (\omega_{rm}
- \mu \alpha_r) (\omega_{sn} - \mu \alpha_s) }{ \omega_{rm} \omega_{sn}
}} s_{rm} s_{sn}.
\end{equation}
As before, the symbol $\approx$ denotes that we have omitted
terms of order $e^{-2 \pi \mu |\alpha_r|}$ (for all $r$).
For $r=s=3$ these are the only nonzero components.
If we define
\begin{equation}
s_{10} = s_{20} = \frac{1}{\sqrt{2}}, \qquad s_{30} = 0,
\end{equation}
then eq.~(\ref{mainresultone}) continues to hold when $m$ is zero
and $n$ is positive (or vice versa).
Finally, if $n=m=0$ we have
\begin{equation}
\overline{N}^{rs}_{00}\approx
 \frac{1}{4 \pi \mu}
\frac{(-1)^{r+s}}{\sqrt{\alpha_r \alpha_s}}, \qquad
\qquad r,s\in \{1,2\},
\end{equation}
\begin{equation}
\overline{N}^{3r}_{00} = \overline{N}^{r3}_{00} = - \sqrt{\alpha_r}, \qquad
r \in \{1,2 \}.
\end{equation}

We remark here that we have been assuming
throughout this work
that $\mu$ is positive,
since it is clear from the analysis of
\cite{Spradlin:2002ar} that only the absolute value of $\mu$
enters into the bosonic matrix elements.
The behavior of various quantities under $\mu
\to -\mu$ was exploited in \cite{Schwarz:2002bc} to derive
several useful identities. Although we have not considered this
involution here, it may be worthwhile to do so.

\section{A Matrix Element}

In this section we use our result to calculate, to
all orders in $\lambda'$, a particular
matrix element of the Hamiltonian which has so far
only been computed to first order in $\lambda'$ in the dual
gauge theory \cite{Pearson:2002zs,Gross:2002mh,Gomis:2002wi}.
We then explain in detail how this matrix element is encoded
in the gauge theory.

\subsection{String field theory}

Consider for $m,n>0$ the three states\footnote{We
caution the reader that the basis of oscillators employed here
and in \cite{Spradlin:2002ar,Spradlin:2002rv}
differs from that used by \cite{Berenstein:2002jq} and
most gauge theory papers by the transformation
$a^{\rm BMN}_n = \frac{1}{\sqrt{2}} (a_{|n|} - i~{\rm sign}(n)
a_{-|n|})$ for $n \ne 0$.}
\begin{eqnarray}
\label{ourstates}
\langle 1 |  &=& {\textstyle{\frac{1}{2}}} \langle 0|
 (a_{m}^i - i a_{-m}^i)(a_{m}^j + i a_{-m}^j),\cr
\langle 2 | &=&  \langle 0|,\cr
\langle 3| &=& {\textstyle{ \frac{1}{2} }}
\langle 0|  (a_{n}^i - i a_{-n}^i)(a_{n}^j +
i a_{-n}^j),
\end{eqnarray}
where $i$ and $j$ are SO(4) indices.
The two-impurity states $\langle 1|$ and $\langle 3|$ decompose
into the ${\bf 1}$, ${\bf 6}$ and ${\bf 9}$ representations
of SO(4).  For definiteness, we fix $i \ne j$, and we could
choose to symmetrize or antisymmetrize in $ij$ at the end of the calculation.
Actually, it turns out that the matrix element vanishes when either
two-impurity state is in the ${\bf 6}$.

We have not discussed the prefactor in this paper, but it has
been shown in \cite{Spradlin:2002rv,Pearson:2002zs} that for states of
the form~(\ref{ourstates}) (in particular, for states with
no fermionic excitations), the three-string coupling
$|H\rangle$ 
in the Hamiltonian is given effectively
by
\begin{equation}
|H \rangle = v(\mu,y) {\cal{P}}|V_B\rangle,
\end{equation}
where
$|V_B\rangle$ was defined in eq.~(\ref{vbdef}) and we
have defined\footnote{The apparent discrepancy between this formula
and the one given in \cite{Pearson:2002zs} is entirely
due to the change of basis in footnote 1.}
\begin{equation}
{\cal{P}} = \frac{\alpha}{2}
\sum_{r=1}^3  \left[
\sum_{m=1}^\infty
\frac{\omega_{rm} }{ \mu \alpha_r} e(m)
a_{rm}^\dagger a_{rm}^{}
\right],
\qquad
e(m) =
\begin{cases}
1 & m \ge 0,\cr
-1 & m < 0,
\end{cases}
\end{equation}
where the SO(4) index (which we have suppressed) is
contracted
between
$a^\dagger$ and $a$.
Finally, the function $v(\mu,y)$ is a measure factor which has not
yet been determined.  It arises from the path integral which defines
the cubic string vertex.  In flat space, the function can be
determined by Lorentz invariance (it is 1 in the supersymmetric theory,
and some very complicated function of $y$ in the bosonic theory).
The plane wave superalgebra does not have enough
generators to fix this overall factor, although comparison with
gauge theory requires that $v(\mu,y) \to 1$ for large $\mu$.
The corresponding factor in the supergravity vertex
in the plane wave background
has been determined to be a constant \cite{Kiem:2002xn}.

For the states~(\ref{ourstates}) we find
the matrix element
\begin{equation}
H_{nmy} \equiv
\langle 1 | \langle 2 | \langle 3| H \rangle =
v(\mu,y)
\frac{1}{4} \frac{\alpha}{2}
\left(\frac{\omega_{1m} }{ \mu \alpha_1}
+ \frac{\omega_{3n} }{ \mu \alpha_3}\right) \left[
2 \left(\overline{N}^{13}_{mn}
\right)^2
- 2 \left(\overline{N}^{13}_{-m,-n})
\right)^2 \right].
\end{equation}
Using eqs.~(\ref{mainresultone}) and~(\ref{mainresulttwo}), we
find after dramatic cancellation the simple result
\begin{equation}
\label{finalH}
H_{nmy} \approx \frac{1}{2 \mu^2} (1-y) \frac{\sin^2(\pi n y) }{\pi^2}
\left[
\frac{v(\mu,y) \mu^2 y }{ \omega_{1m} \omega_{3n}}\right].
\end{equation}
The quantity in brackets is equal to 1 at leading order for large $\mu$,
reproducing the result presented 
in \cite{Pearson:2002zs}.
As is now standard in the literature, it should be understood
that the cubic interaction $H$ considered here enters
the full Hamiltonian with a coefficient equal to the
effective string coupling, $g_2 = 4 \pi g_s \mu^2$.

\subsection{Relation to gauge theory}

The result~(\ref{finalH}) provides a concrete 
all-loop prediction for the gauge theory, which we now explain.
Consider (in the large $J$ limit) the normalized BMN
\cite{Berenstein:2002jq,Beisert:2002bb,Gross:2002mh,
Constable:2002vq}
operators
\begin{eqnarray}
\label{operators}
{{O}}_{n}&=&\frac{1}{ \sqrt{J N^{J+2}}}
\sum_{k=0}^J e^{2 \pi i  k n/J} {\rm Tr}(\phi^i Z^k
\phi^j Z^{J-k}),\cr
{{T}}_{n}^{y}&=&
\frac{1 }{ \sqrt{J y (1-y)}}
\frac{1 }{ \sqrt{J N^{J+2}}}
\sum_{k=0}^{Jy} e^{2 \pi i  k n y/J}
:\!
{\rm Tr}(\phi^i Z^k
\phi^j Z^{Jy- k}) {\rm Tr}(Z^{(1-y) J})\!:,\cr
{{T}}^y &=&
\frac{1 }{ \sqrt{N^{J+2}}}
:\! {\rm Tr}(\phi^i Z^{(1-y) J}) {\rm Tr}(\phi^j Z^{y J})\!:,
\end{eqnarray}
where $i \ne j \in \{1,2,3,4\}$ are SO(4) indices (see
the discussion below eq.~(\ref{ourstates})) labelling
four of the six scalar
fields of ${\cal{N}}=4$ SU(N) gauge theory,
and $Z = \frac{1 }{ \sqrt{2}} (\phi^5 + i \phi^6)$.

At zero string coupling ($g_2 = 0$),
the single- (double-) trace operators
${{O}}$ (${{T}}$) defined in eq.~(\ref{operators})
correspond respectively to one- (two-) string states
and have definite conformal dimensions
\begin{equation}
\Delta_n = J + 2 \sqrt{1 + \lambda' n^2}, \qquad
\Delta_n^y = J + 2 \sqrt{1 + \lambda' n^2/y^2}, \qquad
\Delta^y = J + 2.
\end{equation}
For finite $g_2$,
these operators mix \cite{Beisert:2002bb,Gross:2002mh,
Constable:2002vq}, and
the state-operator correspondence has been worked out to order $g_2$
in \cite{Pearson:2002zs,
Gross:2002mh} (see also \cite{Gomis:2002wi}).
The operators which correspond to the desired one- and two-string
states are
\begin{eqnarray}
\label{stringbasis}
\widetilde{{{O}}}_{p}
&=& {{O}}_p - \frac{g_2 }{ 2} \sum_{k=-\infty}^\infty \int_0^1 dy \ 
C_{pky}\  {{T}}_{k}^{y} - \frac{g_2 }{ 2} \int_0^1 dy\  C_{py} {{T}}^y + 
{\cal{O}}(g_2^2), \cr
\widetilde{T}{}_k^y &=&  {{T}}_{k}^y
- \frac{g_2 }{ 2} \sum_{p=-\infty}^\infty C_{pky}
{{O}}_p + g_2({\rm triple~trace}) + {\cal{O}}(g_2^2),
\end{eqnarray}
where
\begin{equation}
C_{pky} = \sqrt{ \frac{1-y }{ J y} } \frac{\sin^2(\pi p y) }{ \pi^2 (p-k/y)^2},
\qquad C_{py} = - \frac{\sin^2(\pi p y)}{\sqrt{J} \pi^2 p^2}.
\end{equation}
The triple-trace operators in eq.~(\ref{stringbasis})
will not be important for this calculation.
The anomalous dimension matrix element between the single-string
state $\widetilde{O}$ and the two-string state $\widetilde{T}$
is read off from the two-point function
\begin{equation}
(2 \pi x)^{\Delta_n + \Delta_m^y}
\langle \overline{\widetilde{{{O}}}}_n(x)
{\widetilde{{T}}_{m}^y}(0)
\rangle = - g_2
h_{mny}
\ln (x \Lambda)^2.
\end{equation}
The prediction from~(\ref{finalH}), when expressed in
gauge theory variables, is simply
\begin{equation}
\label{prediction}
h_{mny} = \frac{H_{mny}
}{ \sqrt{J y(1-y)}} \approx \frac{\lambda' }{ 2}
\sqrt{ \frac{1 - y }{ J y}} \frac{\sin^2(\pi n y) }{ \pi^2}
\left[ 1 + \lambda' n^2 \right]^{-1/2} \left[1 + \lambda' m^2/y^2
\right]^{-1/2},
\end{equation}
up to the overall function $v(\mu,y)$  discussed in the previous
subsection.
The leading
$\lambda'$ term in this
result agrees with the one-loop field theory calculations
of \cite{Chu:2002pd,Beisert:2002bb,Constable:2002vq}
when the appropriate operator redefinition in eq.~(\ref{stringbasis})
is taken into account.
Note that since eq.~(\ref{prediction}) is already proportional
to $\lambda'$, probing the subleading terms in this expression
would require a two-loop gauge theory calculation, which has
not yet been reported in the literature.

\section{Conclusion and Discussion}

The primary results of this paper are twofold.  First, we have
shown through a quite intricate analysis that it is possible
to determine all Neumann coefficients in the plane wave background
exactly in terms of a single function $F(\mu,y)$,
for any value of the mass parameter $\mu$, and we have provided
an explicit formula for $F$ in eq.~(\ref{Fresult}).
Secondly, 
we have investigated the large $\mu$ behavior of these
coefficients and presented
simple formulas in section 5.6 which give the Neumann matrices
to all orders in a $1/\mu$ expansion.
Note that although we have not discussed the matrix elements of
the prefactor (nor the fermionic Neumann matrices)
in detail, they are very easily obtained  from the Neumann vectors
(\ref{finalnv})
using the results of \cite{Spradlin:2002ar,
Spradlin:2002rv,Pankiewicz:2002gs,Pankiewicz:2002tg}.

Perhaps the most remarkable fact about our results is that
although the Neumann matrices are very complicated functions of $\mu$,
the final expressions (\ref{mainresultone}) and (\ref{mainresulttwo})
for the $1/\mu$ expansion
are very simple.
Ours was a long (and perhaps circuitous) road, and although
we have derived a number of nontrivial identities along the way,
one cannot help but wonder whether there is a more direct
path which yields the same final result.
In particular, our prediction (\ref{prediction}) for a particular
gauge theory calculation is so simple that it cries out for
explanation by some clever argument, perhaps along the lines
of \cite{Santambrogio:2002sb}.

The origin of this simplicity is the fact that the function
$F(\mu,y)$ behaves for large $\mu$ like
\begin{equation}
\label{fexp}
F(\mu,y) = - \ln[4 \pi \mu y(1-y)] + {\cal{O}}(e^{-2 \pi \mu},
e^{-2 \pi \mu y}, e^{-2 \pi \mu (1-y)}),
\end{equation}
with no perturbative corrections (i.e., inverse powers of $\mu$).
This remarkable fact, which we have proven
in two different ways (in appendices C and D),
is reminiscent of various non-renormalization
theorems.
Indeed, although the BMN operators in eq.~(\ref{operators})
are non-BPS in general, their anomalous
dimensions 
\begin{equation}
\label{adim}
\Delta({{O}}_n) -J -2 =  2 \sqrt{1 + \lambda' n^2} -2, \qquad
\lambda' = \frac{ g_{\rm YM}^2 N }{ J^2}
\end{equation}
are nevertheless finite in the limit of large 't Hooft
coupling, provided that $J$ is simultaneously taken to infinity
so that $\lambda'$ remains finite.
This suggests that there should be some residual `effective
supersymmetry' protecting these operators and their interactions.

We have not worked out explicit formulas for the non-perturbative
terms in eq.~(\ref{fexp}), although these could in principle
be obtained by extending the analysis of the appendices.
Finally, the presence of fractional powers of $\lambda'$
in certain string field theory observables
has been
noted in \cite{Spradlin:2002rv,Klebanov:2002mp}.
These surprising powers appear in matrix elements when
the total number of `impurities' is not conserved.  For example,
a matrix element in which two impurities are created or destroyed
is in general a factor of $\mu = 1/\sqrt{\lambda'}$ larger
than a similar matrix element with conserved impurity number.
Although this seems enigmatic from the dual gauge theory
point of view, it has been stressed \cite{Klebanov:2002mp,Vaman:2002ka}
that there is no reason why the large $J$ limit of the $\lambda$
expansion has to agree, order by order in $\lambda$,
with the $\lambda'$ expansion --- especially in light of
the fact that the BMN limit requires taking $\lambda \to \infty$, which
need not be a trivial extrapolation.
For the anomalous dimensions eq.~(\ref{adim}) it seems to work,
but in order to reproduce impurity non-conserving interactions
on the gauge theory side one might
have to sum the $\lambda$ expansion to all orders and then take
the large $J$ limit.

\section*{Acknowledgments}

MS and AV are grateful to I. Klebanov for collaboration
in the early stages of their work on this project, and to D. Freedman
and R. Roiban
for helpful discussions.  The hospitality
of Princeton University (AV) and the Aspen Center
for Physics (JHS, MS and AV), where portions
of this work were carried out, is also appreciated.
This research was supported in part by the
U.S. Dept.~of Energy under Grants No.~DE-FG02-95ER40893 (YHH),
DE-FG03-92-ER40701 (JHS) and DE-FG02-91ER40671 (MS),
and by the National Science Foundation under
Grant No.~PHY99-07949 (AV).

\appendix

\section{Basic Identities}

The infinite matrices $A^{(r)}_{mn}$, $C$ and the infinite vector $B_m$
satisfy a number of useful relations which we record here:
\begin{equation}
A^{(r) {\rm T}} C A^{(s)} = \frac{1}{\alpha_r}\, C \delta^{rs}, \qquad r,s
\in \{1,2\},
\label{usrelone}
\end{equation}
\begin{equation}
A^{(r) {\rm T}} C^{-1} A^{(s)} =  {\alpha_r }\, C^{-1} \delta^{rs}, \qquad
r,s \in \{1,2\},
\label{usreltwo}
\end{equation}
\begin{equation}
A^{(r) {\rm T}} C B = 0, \qquad r \in \{1,2 \},
\end{equation}
and
\begin{equation}
B^{\rm T} C B = \frac{2}{y(1-y)},
\end{equation}
where we have used $\alpha_3 = -1$,
Some additional useful identities are
\begin{equation}
\sum_{r=1}^3 \frac{1}{ \alpha_r} A^{(r)} C A^{(r){\rm T}} =0,
\end{equation}
\begin{equation}\label{AAT2}
\sum_{r=1}^3 \alpha_r A^{(r)} C^{-1} A^{(r){\rm T}} = \frac{\alpha}{2} B
B^{\rm T},
\end{equation}
where, as before,
\begin{equation}
\alpha = \alpha_1 \alpha_2 \alpha_3 = -y(1-y).
\end{equation}

\section{Derivation of the Differential Equation}

In this appendix we fill in the details of the derivation in section 3,
making frequent use of the identities presented in appendix A.

We start by deriving eq.~(\ref{recast}).
Using eq.~(\ref{usrelone}) to simplify 
$\Gamma_+ C A^{(r)}$
gives the identity
\begin{equation}
\label{ba}
\Gamma_+^{-1} C_3 A^{(r)}=C A^{(r)}-\frac{1}{\alpha_r} \Gamma^{-1}_+ A^{(r)}
C_r, \qquad r \in \{1,2\}.
\end{equation}
It follows from this that
\begin{equation}
\label{bb}
Y^{\rm T} C U_3^{-1} A^{(r)} = - \frac{1}{\alpha_r} Y^{\rm T} A^{(r)}
U_r^{-1} C,
\qquad r \in \{1,2\}.
\end{equation}
Multiplying eq.~(\ref{symformula}) on the right by $A^{(r)}$ and
using eqs.~(\ref{ba}) and~(\ref{bb}) then gives
\begin{equation}
\Gamma_+^{-1} A^{(r)} C_r - \alpha_r C_3 \Gamma_+^{-1} A^{(r)} =
- \frac{1}{2} \frac{\alpha}{1 + \mu \alpha k} C U_3^{-1} Y  Y^{\rm T}
A^{(r)} U_r^{-1} C, \qquad r \in \{1,2\}.
\end{equation}
Comparing this to eq.~(\ref{symformula}), we can 
write the following formula for all $r$,
\begin{equation}
\Gamma_+^{-1} A^{(r)} C_r - \alpha_r C_3 \Gamma_+^{-1} A^{(r)} =
C \delta_{r3}- \frac{1}{2} \frac{\alpha}{1+\mu \alpha k}
C U_3^{-1} Y Y^{\rm T} A^{(r)} C U_r^{-1}.
\end{equation}
Multiplying this further by $B^{\rm T} C_3^{-1}$ on the left
and by $\alpha_r C^{-1} C_r^{-1} A^{(r) {\rm T}}$ on the right,
then summing over $r$, gives
\begin{equation}
\label{bc}
\frac{1}{2} \alpha k_1 B^{\rm T}
- Y^{\rm T} N =
- B^{\rm T} C_3^{-2}
- \frac{1}{ 2} \frac{\alpha }{ 1+\mu \alpha k}
(k - \mu k_1)
(\frac{1}{2} \alpha k B^{\rm T} + \mu Y^{\rm T} N ).
\end{equation}
The transpose of eq.~(\ref{bc}) is a linear equation for $NY$ whose solution
is~(\ref{recast}).

Next we obtain eq.~(\ref{Yeq}).
We start by 
using a variant of eq.~(\ref{symformula}) 
to obtain an identity for
$[C_3^{-2}, \Gamma_+^{-1} ] = [C_3^{-1},
\{ \Gamma_+^{-1}, C_3^{-1} \}]$ which yields
\begin{equation}
[C_3^{-2},\Gamma^{-1}_+] B=
- \frac{1}{2}
\frac{\alpha }{ 1+ \mu \alpha k}
[(k-\mu k_1) (C_3^{-1}-\mu C_3^{-2})-
(k_1-\mu k_2)(1-\mu C_3^{-1})
]Y
\end{equation}
and hence
\begin{equation}
\label{arrive}
\Gamma_+^{-1} C_3^{-2} B = \frac{1}{ 2} \frac{\alpha}{1 + \mu \alpha k}
\left[ (k_2 \mu - k_1) + (k - \mu^2 k_2) C_3^{-1} + (\frac{2}{\alpha}
+ k \mu + k_1 \mu^2)
C_3^{-2}  \right]Y.
\end{equation}
Then we can substitute eq.~(\ref{recast}) into
eq.~(\ref{follows}) and use eq.~(\ref{arrive}) to arrive
at eq.~(\ref{Yeq}).

\section{On an Integral Transform}

In this appendix we show that the general solution
of the Fredholm integral equation of the first kind
\begin{equation}
\label{inteq}
\int_0^\infty dx
\frac{f(x)}{(x^2 + z^2)^{3/2}} = g(z)
\end{equation}
is
\begin{equation}
\label{gensoln}
f(x) = i \frac{x^2}{\pi} \int_0^{\pi/2} \left[ g(i x \cos \theta)
- g (-i x \cos \theta)\right] \cos \theta d \theta.
\end{equation}
For the amusement of the reader we present two derivations
of this result.  The first is a set of manipulations
that indicate how we arrived at the solution, while the second
is a more rigorous proof of the theorem as stated in the text.

\subsection{An elementary manipulation}

As integral equations of the first kind are notoriously difficult to
solve, let us attempt to circumvent the problem. The method used here
does not seem to be in the canonical texts and is hoped to be of some
use.

Consider the following integral identity \cite{GR}:
\begin{equation}
\label{Jid1}
\int_0^\infty \frac{xJ_0(x j) dx}{(x^2 + z^2)^{3/2}} =
\frac{e^{-zj}}{z}, \qquad j>0,~{\rm Re}(z)>0.
\end{equation}
Next suppose that $zg(z)$ can be expanded as
a power series in $e^{-z} -1$:
\begin{equation}
\label{gexpand}
z g(z) = \sum_{j=0}^\infty a_j (e^{-z}-1)^j 
= \sum_{j=0}^\infty a_j \sum_{k=0}^j
\binom{j}{k}
(-1)^{j-k} e^{-zk}.
\end{equation}
Therefore by eq.~(\ref{Jid1}), we have
\begin{eqnarray*}
g(z)
&=& \sum_{j=0}^\infty a_j \sum_{k=0}^j \binom{j}{k}(-1)^{j-k}
        \frac{e^{-zk}}{z} \\
&=& \sum_{j=0}^\infty a_j \sum_{k=0}^j \binom{j}{k}(-1)^{j-k}
        \int_0^\infty \frac{xJ_0(x k) dx}{(x^2 + z^2)^{3/2}},
\end{eqnarray*}
from which we can directly read off the solution
\begin{equation}
f(x) = x \sum_{j=0}^\infty a_j \sum_{k=0}^j \binom{j}{k}(-1)^{j-k}
J_0(xk).
\end{equation}

This formidable sum can actually be performed due to the identity
\cite{GR}
\begin{equation}
\label{Jid2}
J_0(k x) = \frac{1}{\pi} \int_0^\pi e^{i k x \cos \theta} d \theta,
\end{equation}
from which we obtain
\begin{eqnarray}
\label{solution}
f(x)
&=& \frac{x}{\pi} \sum_{j=0}^\infty a_j
        \sum_{k=0}^j \binom{j}{k}(-1)^{j-k} \int_0^\pi (e^{i x \cos
        \theta})^k d \theta  \\
\nonumber
&=& \frac{x}{\pi}  \int_0^\pi \sum_{j=0}^\infty a_j (e^{i \cos x
        \theta} - 1) ^j d\theta \\ \nonumber
&=& \frac{-i x^2}{\pi} \int_0^\pi d\theta \cos \theta 
        \ g(-i x \cos \theta).
\end{eqnarray}
Therefore all dependence on the coefficients $a_j$ drops out and we
have a generalized Fourier transform of rather simple form.

The main shortcoming of this derivation is that it assumes the existence
of a rather peculiar expansion, which might not be necessary. The 
alternative approach presented in the remainder of this appendix is
clearer in this regard.

\subsection{Representation of a delta-function}

Here we consider the integral
\begin{equation}
\label{deltaone}
\chi(y,y') =
\frac{\sqrt{y}}{4 \pi i} \int_{\cal C} \frac{dw}{\sqrt{1 + w}} \frac{1}{
(w y + y')^{3/2}},
\end{equation}
where $y,y'>0$ and ${\cal{C}}$ is the contour shown in
fig.~1.  For $y'<y$ the singularity structure is shown in
fig.~1(a), and the contour can be pushed off to infinity, giving
zero for the integral.
For $y'>y$, the contour encloses no singularity (fig.~1(b)), so
the result is again zero.

\begin{figure}
\centering{\psfig{figure=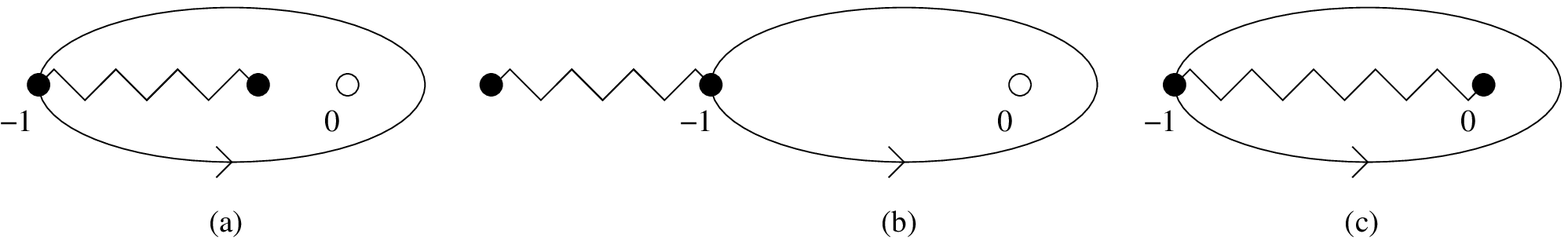,width=5.0in}}
\caption{The contour ${\cal{C}}$ for appendices C and D.
In (a) and (b) we show the singularity structure of eq.~(\ref{deltaone}),
with branch points at $-1$ and $-y'/y$.  In (c) we show the
structure of eqs.~(\ref{onetwoseven}) (for integer $k$)
and~(\ref{onefourone}).}
\end{figure}

Since
$\chi(y,y')$ vanishes for $y'\ne y$,
let us check whether it is a delta-function (and in particular,
that it does not involve any derivatives of delta-functions)
by integrating
it against the test function $e^{-t y'}$:
\begin{equation}
\label{remain}
\int_0^\infty dy'\ \chi(y,y') e^{-t y'}
=  \frac{1}{2 \pi i} \int_{\cal{C}} \frac{dw}{\sqrt{1 + w} \sqrt{w}}
\left[1 + e^{t w y} \sqrt{\pi t w y} (\Phi(\sqrt{t w y}) - 1)\right],
\end{equation}
where the result of the $y'$ integral involves the error function
\begin{equation}
\Phi(x) = \frac{2}{\sqrt{\pi}} \int_0^x e^{-t^2} dt.
\end{equation}
Next we make use of the elementary integral
\begin{equation}
\label{onetwoseven}
\frac{1}{2 \pi i} \int_{\cal{C}} \frac{dw}{\sqrt{1 + w} \sqrt{w}}
w^k = \binom{- \frac{1}{2}}{ k}.
\end{equation}
In particular, note that
this is zero when $k$ is a positive half-integer
since the branch cut then runs from $-1$ to $-\infty$, so there is
no singularity within the contour.
The term in eq.~(\ref{remain}) proportional to ``$-1$'' therefore
integrates to zero, since it contains only half-integer powers
of $w$.  For the remaining terms we use the expansion
\begin{equation}
\label{expansion}
1 + e^x \sqrt{\pi x} \Phi(\sqrt{x}) = \sum_{k=0}^\infty
\frac{2^k x^k}{(2k-1)!!}.
\end{equation}
Note that $(-1)!!=1$.
Combining eqs.~(\ref{expansion}) and~(\ref{onetwoseven})
leads to
\begin{equation}
\label{results}
\int_0^\infty dy' \ \chi(y,y') e^{-t y'} = \sum_{k=0}^\infty
\frac{2^k (t y)^k }{(2k-1)!!} \binom{- \frac{1}{2}}{k}
= \sum_{k=0}^\infty \frac{ (-1)^k}{ k!} (ty)^k = e^{-ty},
\end{equation}
where we have used the identity
$(2k-1)!! = (-2)^k k! \binom{- \frac{1 }{ 2}}{ k}$.
If $\chi$ contained any terms proportional to derivatives
of delta-functions, we would have obtained a polynomial in
$t$ times $e^{-ty}$ in eq.~(\ref{results}).
Since this did not happen, we have proven that
\begin{equation}
\label{deltafun}
\chi(y,y') = \frac{\sqrt{y}
}{4 \pi i} \int_{\cal C} \frac{dw }{\sqrt{1 + w}} \frac{1}{
(w y + y')^{3/2}}
= \delta(y-y'), \qquad y,y'>0.
\end{equation}
More precisely, we have proved that $\int \chi(y, y') f(y') dy' = f(y)$
for any $f$ that can be written as a convergent Laplace transform. It is not
excluded that an even weaker condition would suffice.

\subsection{Formal proof}

To verify our solution
let us substitute $f(x)$ from eq.~(\ref{gensoln}) into the
integral equation (\ref{inteq}) 
and replace the integration variable $x$ by $t=
x \cos \theta $. This gives
\begin{equation}
\label{bd}
-\frac{i}{ \pi} \int_0^\infty dt \int_0^{\pi/2}
 d \theta \frac{ t^2 \left[ g(-i t) - g(i t) \right] \cos \theta}{
(t^2 + z^2 \cos^2 \theta)^{3/2}} =
-\frac{i}{\pi} \int_0^\infty dt  \frac{t [ g(-it) - g(it)]
 }{(t^2 + z^2)},
\end{equation}
where we have used
\begin{equation}
\int_0^{\pi/2} \frac{\cos \theta d \theta }{ (t^2 + z^2
\cos^2 \theta)^{3/2}} = \frac{1 }{ t(t^2 + z^2)}.
\end{equation}
We can now recast the remaining integral in eq.~(\ref{bd})
as a contour integral with the
contour ${\cal C}'$ enclosing the positive $u$ axis. This gives
\begin{equation}
  \frac{1}{2\pi i} \int_{{\cal C}'} du \frac{ g(-i\sqrt{u})
 }{ (u + z^2)} =g(z),
\end{equation}
where we have used the assumed analytic and asymptotic 
properties of $g$, which are just
what are needed to deform the contour into one that encircles the
pole at $-z^2$, thereby reducing the integral to exactly $g(z)$. 

We have proven that the function $f(x)$ given in eq.~(\ref{gensoln}),
when substituted into eq.~(\ref{inteq}), indeed gives $g(z)$.
It remains to show that the result is unique, in other words that
when $g(z)$ from eq.~(\ref{inteq}) is substituted into eq.~(\ref{gensoln}),
we recover $f(x)$.
To this end, we define first
\begin{equation}
f(x) = x \tilde{f}(x^2),
\end{equation}
and let $v' = x^2$ so that the integral
equation~(\ref{inteq}) becomes
\begin{equation}
\label{tildef}
\frac{1}{2} \int_0^\infty dv' (z^2 + v')^{-3/2} \tilde{f}(v') = g(z).
\end{equation}
Our proposed solution is, in terms of $\tilde{f}(v)$,
\begin{equation}
\tilde{f}(v) = -\frac{i \sqrt{v}}{\pi} \int_0^{\pi}
g(-i \sqrt{v} \cos \theta) \cos \theta d \theta.
\end{equation}
Letting $\sqrt{w} = - i \cos\theta$ gives
\begin{equation}
\label{bg}
\tilde{f}(v) = \frac{\sqrt{v} }{ 2 \pi i} \int_{\cal{C}}
\frac{dw }{ \sqrt{1 + w}} g(\sqrt{w v}),
\end{equation}
where the contour ${\cal{C}}$ is as in fig.~1.
Substituting eq.~(\ref{inteq}) for $g(z)$ and using the
delta-function representation eq.~(\ref{deltafun}) establishes
eq.~(\ref{bg}) and hence completes the proof.

\section{Asymptotic Behavior of $F(\mu,y)$}

The analysis of appendix C was completely general, and we only used
rather weak assumptions about the form of $g(z)$.  Here we use
the contour integral techniques above to study the particular
case of interest, with $g(z)$ given by $G(z,y)$ from eq.~(\ref{Gdef}).
Recall Stirling's asymptotic expansion
\begin{equation}
\label{stirling}
\ln
\Gamma(1 + z) \approx (z + {\textstyle{\frac{1}{2}}}) \ln z - z + \frac{1}{2}
\ln 2 \pi + \sum_{n=0}^\infty \frac{c_{n} }{ z^{2n+1}}.
\end{equation}
The coefficients $c_{n}$ may be expressed in terms of Bernoulli numbers,
but they will turn out to be irrelevant.  The crucial fact is
that only odd inverse powers of $z$ appear in eq.~(\ref{stirling}).
It follows that
\begin{equation}
\label{Gasym}
G(z,y) \approx  - \frac{1 }{ z^2} \ln (2 \pi z y(1-y))
+ \sum_{n=1}^\infty\frac{b_n(y) }{ z^{2n+1}}.
\end{equation}
Now letting $\sqrt{w} = - i \cos \theta$ in~(\ref{Fis}), as
in the previous subsection, gives
\begin{equation}
F(\mu,y) = \frac{\mu^2 }{ 2 \pi i} \int_{\cal{C}} \frac{dw }{
\sqrt{1+w}}  G(\mu \sqrt{w}, y).
\end{equation}
Consider first the inverse odd powers in 
eq.~(\ref{Gasym}).  They lead to integrals of the form
\begin{equation}
\label{onefourone}
\int_{\cal{C}} \frac{dw }{ \sqrt{1 + w}} \frac{1}{w^{n+1/2}}
\end{equation}
for $n$ a positive integer.
The branch cut is now as shown in fig.~(1c), so the contour
may be pulled off to infinity, giving zero.
Therefore only the first term in eq.~(\ref{Gasym}) contributes, so
\begin{eqnarray}
F(\mu,y) &\approx& - \frac{1 }{2 \pi i}
\int_{\cal{C}} \frac{dw }{ w \sqrt{1 + w}} \ln(2 \pi \mu \sqrt{w} y(1-y))
\cr
&\approx& - 
\ln(2 \pi \mu y(1-y)) \frac{1}{ 2 \pi i}
\int_{\cal{C}} \frac{dw }{ w \sqrt{1 + w}}
- \frac{1 }{ 4 \pi i} \int_{\cal{C}} \frac{\ln w }{ w\sqrt{1 + w}} dw.
\end{eqnarray}
The first integral gives 1 since it just picks off the pole at $w=0$.
It is an interesting though straightforward exercise to check that
the second term gives $-\ln 2$.  This completes
the direct proof of eq.~(\ref{Fexpansion}).

\section{Advanced Sums}

Let us review an elementary trick which can be used to evaluate
certain
infinite sums.
Consider first the sum
\begin{equation}
\sum_{n=-\infty}^\infty f(n), \qquad
f(z) = \frac{1 }{ z^2 + x^2 v^2} \frac{1 }{ \sqrt{z^2 + x^2}},
\end{equation}
for $x > 0$ and $v>1$.
This sum can be written as the contour integral
\begin{equation}
\frac{1 }{ 2 \pi i} \int_{\cal{C}}dz\ f(z) \pi \cot \pi z,
\end{equation}
where ${\cal{C}}$ passes from $-\infty - i \epsilon$ to
$+ \infty - i \epsilon$ slightly below the real axis, and then
returns slightly above the real axis (see figure~2).
The contour may be deformed away from the
poles of $f(z) \pi \cot \pi z$ on the
real axis to pick up the other singularities instead.
In this case the only other singularities
are branch cuts from $\pm i x$ to $\pm i \infty$, and poles
at $z = \pm i x v$ sitting on the branch cuts.
Rescaling $z$ and combining the two cuts, we find therefore that
\begin{equation}
\label{sumone}
\sum_{n=-\infty}^\infty \frac{1 }{ n^2 + x^2 v^2} \frac{1 }{ \sqrt{n^2 + x^2}}
= - \frac{2 }{x^2} {\rm P} \!\int_1^\infty
\frac{dz }{ \sqrt{z^2 - 1}}\frac{\coth(\pi x z) }{ z^2 - v^2},
\end{equation}
where P stands for the principal value.
This is an exact formula, but if we
are not concerned with terms vanishing exponentially for large
$x$, then we can set $\coth(\pi x z) = 1$ in the integral, which
is then easily evaluated to give
\begin{equation}
\sum_{n=-\infty}^\infty \frac{1 }{ n^2 + x^2 v^2} \frac{1}{\sqrt{n^2 + x^2}}
\approx \frac{2}{x^2}\frac{ {\rm arc~cosh}(v) }{ v \sqrt{v^2 - 1}}.
\end{equation}
As in section 5, the symbol $\approx$ denotes that we have
dropped terms of order $e^{-2 \pi x}$.
Taking the limit $v \to 1$ from above finally gives the result
\begin{equation}
\label{sumoneb}
\sum_{n=-\infty}^\infty \frac{1 }{(n^2 + x^2)^{3/2}} \approx \frac{2 }{x^2}.
\end{equation}

\begin{figure}
\centering{\psfig{figure=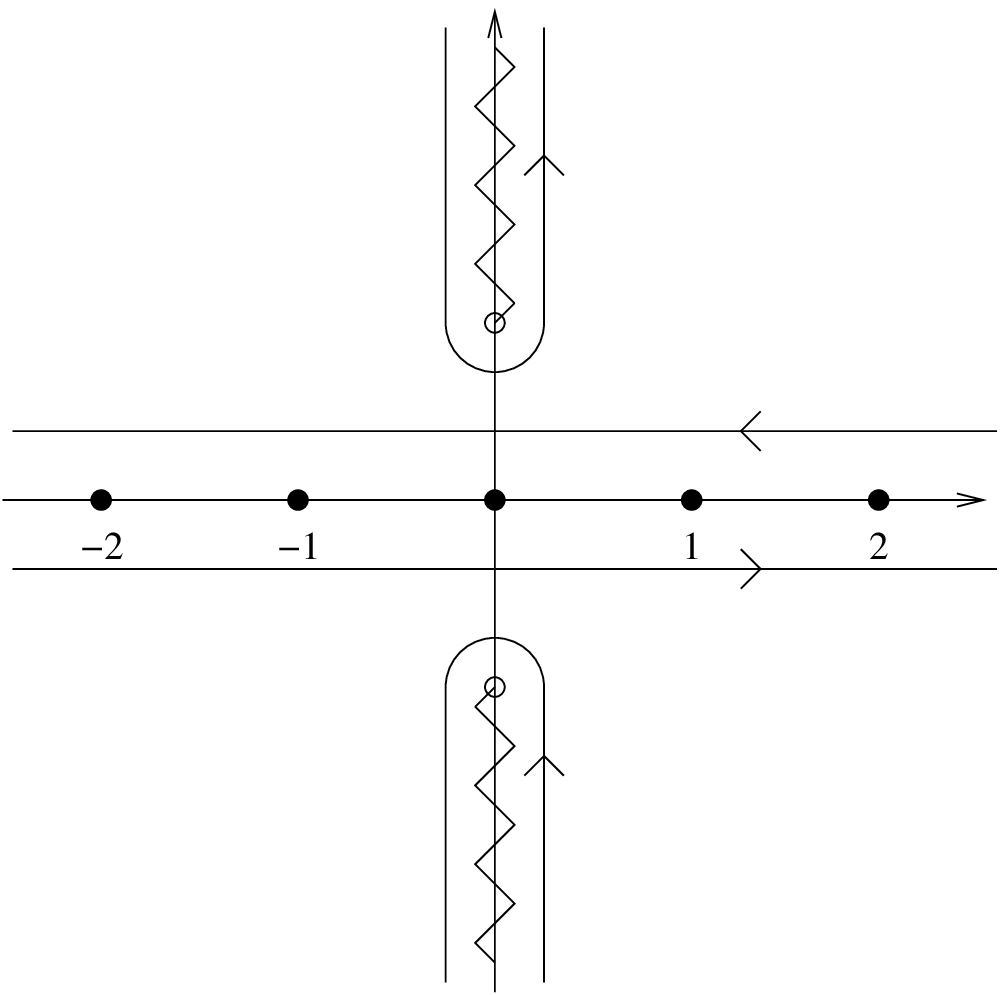,width=2.5in}}
\caption{This figure shows the contours relevant for analyzing
the various sums and integrals in appendix E.  In each case,
the contour is deformed from the one enclosing the real axis to the
one enclosing the branch cuts on the imaginary axis.
Note that the integrands in eqs.~(\ref{sumone}) and~(\ref{sumfive})
have two additional poles on top of the branch cuts, while
the integrand in eq.~(\ref{sumtwo}) has two additional poles on the real
axis.}
\end{figure}

For sums in which $f(z)$ itself contains trigonometric
functions, it is convenient to expand these and
write the sum in the form
\begin{equation}
\sum_{n=-\infty}^\infty [f_1(n) + (-1)^n f_2(n)].
\end{equation}
The term with $f_1(z)$
can be evaluated using the above trick, while the second
term
can be evaluated by deforming the contour
integral of $f_2(z) \pi \csc \pi z$.
Applying this method and taking into account
both the contribution from
the poles at $\pm n/y$ and the discontinuity across the branch cuts
gives
\begin{equation}
\label{sumtwo}
\sum_{m=1}^\infty \frac{\sin^2(\pi m y)}{
m^2 - n^2/y^2} \frac{ \sqrt{x + \sqrt{x^2 + m^2}}
}{ \sqrt{x^2 + m^2}}
= - \frac{1}{2 x^{3/2}} {\rm Re} \int_1^\infty \frac{dz}{\sqrt{z^2 - 1}}
\frac{\sqrt{1 + i \sqrt{z^2 - 1}}}{ z^2 + n^2/(x^2 y^2)} \frac{1}{ F(x,y,z)}
\end{equation}
when $0<y<1$, $x>0$, and $n$ is a non-zero integer.
We have defined
the function
\begin{equation}
F(x,y,z)  = \frac{1 }{ 2} \left[
\coth(\pi x y z) + \coth(\pi x (1-y) z)\right].
\end{equation}
Ignoring exponentially small corrections, we can set $F=1$
and evaluate the integral in eq.~(\ref{sumtwo}) explicitly, arriving at
\begin{equation}
\label{sumthree}
\sum_{m=1}^\infty \frac{ \sin^2(\pi m y) }{ m^2 - n^2/y^2}
\frac{ \sqrt{x + \sqrt{x^2 + m^2}} }{ \sqrt{x^2 + m^2}}
\approx - \frac{\pi y^{3/2} }{ 4  n} \frac{\sqrt{\sqrt{n^2 + x^2 y^2} - x y}
}{ \sqrt{n^2 + x^2 y^2}}.
\end{equation}
The case $n=0$ must be considered separately since the poles
at $m^2 = \pm n/y$ are then lost.  The result in this case is
\begin{equation}
\label{sumfour}
\sum_{m=1}^\infty \frac{\sin^2 (\pi m y) }{ m^2}
\frac{\sqrt{x + \sqrt{x^2 + m^2}} }{\sqrt{x^2 + m^2}}
\approx \frac{\pi^2 y(1-y) }{ \sqrt{2 x}} - \frac{\pi }{ 4 \sqrt{2}
x^{3/2}}.
\end{equation}
A combination of all of the above techniques is needed
to tackle the final sum
\begin{eqnarray}
\label{sumfive}
\sum_{m=1}^\infty \frac{\sin^2(\pi m y)}{ m^2 + x^2 v^2}
\frac{\sqrt{x + \sqrt{x^2 + m^2}} }{ \sqrt{x^2 + m^2}}
&\approx& - \frac{2 }{ x^{3/2}} {\rm Re}\,
{\rm P} \!\int_1^\infty \frac{dz }{ \sqrt{z^2 - 1}}
\frac{\sqrt{1 + i \sqrt{z^2 - 1}} }{ z^2 - v^2}
\cr
&=&
- \frac{\pi }{ 2 x^{3/2}}
\frac{{\rm Im} \sqrt{1 - i \sqrt{v^2 - 1}} }{
v \sqrt{v^2 - 1}},
\end{eqnarray}
which is valid for $v>1$.

\section{Some Exponential Corrections}

In this appendix we derive the exact formula~(\ref{fullF}) which
allows systematic determination of the exponential corrections
to $F(\mu,y)$.
It follows from eq.~(\ref{sumone}) that
\begin{equation}
\phi''(x) = - x \sum_{n=1}^\infty \frac{1}{(x^2 + n^2)^{3/2}}
= \frac{1}{2 x^2} + \frac{1}{x} I(x)
\end{equation}
where
\begin{equation}
I(x) = {\rm P}\int_1^\infty \frac{\coth(\pi x z)}{(z^2-1)^{3/2}}
dz = - \frac{1}{2} \oint_{|z-1| = \epsilon} \frac{\coth(\pi x z)}{
(z^2-1)^{3/2}} dz + \int_{1+\epsilon}^\infty
\frac{\coth(\pi x z)}{(z^2-1)^{3/2}}
dz.
\end{equation}
Integrating the second term by parts
gives a divergent piece that cancels the divergent
piece from the first term, leaving
\begin{equation}
I(x) = - 1 - \pi x \int_1^\infty
\frac{z dz}{\sqrt{z^2-1}}
\frac{1}{\sinh^2(\pi x z)}.
\end{equation}
Therefore
\begin{equation}
\label{phippexact}
\phi''(x) = - \frac{1}{x} + \frac{1}{2 x^2} - \pi \int_1^\infty
\frac{z dz}{\sqrt{z^2-1}} \frac{1}{\sinh^2(\pi x z)}.
\end{equation}
Integrating twice with respect to $x$ as in subsection 5.1 and
using eq.~(\ref{Fresult}) yields
the formulas~(\ref{fullF}) and~(\ref{Jdef}).

\end{document}